\documentclass[12pt]{iopart}

\usepackage{amssymb}
\usepackage{graphicx}
\usepackage{color}
\usepackage[implicit=true,colorlinks=true,linkcolor=blue,
citecolor=blue,urlcolor=blue]{hyperref} 

\newcommand{\dd}{\mathrm{d}}
\newcommand{\ii}{\mathrm{i}}
\newcommand{\ee}{\mathrm{e}}
\newcommand{\eqref}[1]{(\ref{#1})}

\newcommand{\ket}[1]{|#1\rangle}

\include{MyMc}

\begin{document}

\title{A tunable macroscopic quantum system based on two fractional vortices}

\author{D\,M~Heim$^1$, K~Vogel$^1$, W\,P~Schleich$^1$, D~Koelle$^2$, R~Kleiner$^2$ and E~Goldobin$^2$}

\address{$^1$ Institut f{\"u}r Quantenphysik and Center for Integrated Quantum Science and Technology (IQ$^\mathrm{ST}$), Universit{\"a}t Ulm, D-89069 Ulm, Germany}

\address{$^2$ Physikalisches Institut and Center for Collective Quantum Phenomena in LISA$^+$, Universit\"{a}t T\"{u}bingen, D-72076 T\"{u}bingen, Germany}

\ead{dennis.heim@uni-ulm.de}

\begin{abstract}
{{}}{We propose a tunable macroscopic quantum system based on two fractional vortices. Our analysis shows that two coupled fractional vortices pinned at two artificially created $\kappa$ discontinuities of the Josephson phase in a long Josephson junction can reach the quantum regime where coherent quantum oscillations arise. For this purpose we map the dynamics of this system to that of a single particle in a double-well potential. By tuning the $\kappa$ discontinuities with injector currents we are able to control the parameters of the effective double-well potential as well as to prepare a desired state of the fractional vortex molecule. The values of the parameters derived from this model suggest that an experimental realisation of this tunable macroscopic quantum system is possible with today's technology.}
\end{abstract}



\maketitle

\section{Introduction}
\label{Sec:introcuction}

{{}}{What experimental evidence do we have that quantum mechanics is valid at the macroscopic level? This question raised~\cite{Legett:1980} in 1980 by A. J. Leggett has triggered a flood of theoretical work~\cite{leggett:1987:rev,haenggi:1990:rev,zurek:2003:rev} and experiments on a wide variety of quantum systems ranging from photons in cavities~\cite{brune:1996:cavity,deleglise:2008:haroche:nature}, ions in traps~\cite{monroe:1996:wineland:science}, cold atoms~\cite{Andrews:1997:bec} via high-spin molecules~\cite{friedman:1996} to superconducting devices~\cite{clarke:1988}. In the present paper we show that it is possible to tune two fractional Josephson vortices into the quantum regime and obtain in this way a system in which we can observe macroscopic quantum phenomena.}

\subsection{Coherent quantum oscillations}

{{}}According to Leggett~\cite{Legget:2002:TestingTheLimits} the first experiments produced a quantum superposition of macroscopically distinct states. In order to validate this situation experimentally, Leggett proposed~\cite{Legget:2002:TestingTheLimits} a ``real-time'' experiment concerning the two classical ground states $\ket{L}$ and $\ket{R}$ of a double well potential: ``one starts the system off in, say, $\ket{L}$, switches one's measuring apparatus off, then switches it on again at time $t$ and detects whether it is in $\ket{L}$ or $\ket{R}$, and by making repeated runs of this type with varying values of $t$, plots a histogram of the probability $P_L(t)$ that the system is in $\ket{L}$ at time $t$.'' A quantum-mechanical calculation would yield
\begin{equation}
P_L(t) = [1 + \cos ( \Delta_{01} t)]/2
, \label{eq:P_L}
\end{equation}
{{}}{where $\Delta_{01}=(E_1-E_0)/\hbar$ and $E_1-E_0$} is the energy splitting separating the two lowest quantum-mechanical energy levels of the double-well potential.

{
This experiment was realised with superconducting charge~\cite{Nakamura:1999}, phase~\cite{Poletto:2009} and flux~\cite{Fedorov:2011} devices, where quantum oscillations were based on Cooper-pairs, the Josephson phase and the flux generated by a superconducting ring, respectively. But until now, no experiment reported coherent quantum oscillations for Josephson vortices. These are vortices of electric current appearing in superconducting Josephson junctions (JJs). Their macroscopic quantum nature was only demonstrated in escape experiments~\cite{Wallraff:2003:Fluxon:QuTu}, where the macroscopic ground state tunnels out of a potential well.
}

{
In the present paper we propose a ``real-time'' experiment for Josephson vortices based on the experimental setup described in reference~\cite{Kienzle:2012:Molecule:Spectroscopy}. Furthermore, we show how our system can be tuned from a classical regime, where no quantum oscillations are observable, into the quantum regime.

There is also a different concept for observing such vortex oscillations~\cite{price:2010}. It does not involve injector currents which are sources of {{}}{noise} in our device. On the other hand this alternative concept requires a bias current and an external magnetic field which also brings {{}}{fluctuations} into the system. These two components are absent in our setup.
}

\subsection{Long Josephson junctions}

Both approaches are based on long Josephson junctions (LJJs). These devices attracted the attention of researchers during several decades because it is a relatively clean and well-controlled non-linear system, which allows to study fundamental physics. For example, Cherenkov radiation from a fast moving fluxon~\cite{Goldobin:Cherry2} was observed, the ratchet effect~\cite{Carapella:RatchetE:2001,Ustinov:2004:BiHarmDriverRatchet,Knufinke:2012:JVR-loaded} was demonstrated and the Zurek-Kibble scenario of phase transitions~\cite{Monaco:2006:Zurek-Kobble.New} was tested. Furthermore, LJJs have applications, \ie, as flux flow oscillator in submillimeter receivers~\cite{Koshelets:2007:IntRec4TELIS}. New possibilities have opened with the advent of $\pi$ Josephson junctions and technologies allowing to combine conventional 0 and $\pi$ junctions in one device~\cite{Weides:SF:SFS,sickinger:2012}.

In a one-dimensional $0$-$\pi$ LJJ vortices carrying half of the magnetic flux quantum $\Phi_0\approx 2.07\times10^{-15}\units{Wb}$ appear spontaneously at the boundaries between $0$ and $\pi$ regions~\cite{Bulaevskii:0-pi-LJJ,buzdin:1982}. These semifluxons~\cite{Xu:SF-shape,Goldobin:SF-Shape}
have a degenerate ground state of either positive or negative polarity. The orientation \state{s} or \state{a} depends on the direction of supercurrent circulating around the 0-$\pi$ boundary. The classical behaviour of semifluxons has been studied extensively theoretically~\cite{Bulaevskii:0-pi-LJJ,Xu:SF-shape,Goldobin:SF-Shape,Susanto:SF-gamma_c} 
and experimentally~\cite{Kirtley:SF:HTSGB,Hilgenkamp:zigzag:SF,rocca:2005,cedergren:2010}. 

One can also construct ``molecules'' consisting of two or more semifluxons, for example, a molecule with antiferromagnetically (AFM) arranged semifluxons and antisemifluxons situated at a distance $a$ from each other in a $0$-$\pi$-$0$ LJJ. Such a molecule exhibits two degenerate ground states \state{sa} or \state{as}. One can switch the molecule between the \state{sa} and \state{as} states by applying a small uniform bias current to the LJJ~\cite{kato:1997,Goldobin:SF-ReArrange,Dewes:2008:ReArrangeE}.

Such an AFM semifluxon molecule was suggested to observe quantum {{}}{oscillations} of semifluxons~\cite{Goldobin:2005:MQC-2SFs}.
The low energy dynamics of the system was reduced to the dynamics of a point-like particle in a double-well potential and the parameters of this potential and the mass of the particle were related to the parameters of the 0-$\pi$-0 LJJ. The parameters for which quantum effects emerge were estimated. However, in such a double-well potential the height of the barrier depends only on \emph{geometrical} parameters of the 0-$\pi$-0 LJJ, in particular, on the length $a$ of the $\pi$ part, which makes it impossible to tune the barrier during experiment.

In a conventional LJJ with the Josephson phase $\phi$ and the first Josephson relation $j_s=j_c\sin\phi$ one can create artificially an \emph{arbitrary}, \emph{electronically tunable} $\kappa$ discontinuity of the phase $\phi$, so that $\phi(x)=\mu(x)+\kappa\Heavyside(x)$ and $\mu(x)$ is a smooth function without discontinuities and $\Heavyside(x)$ is a Heaviside step function. For $\kappa=\pi$ this junction is equivalent to the 0-$\pi$ LJJ described above. Such a device was proposed and successfully tested~\cite{Ustinov:2002:ALJJ:InsFluxon,Gaber:2005:NonIdealInj2,Goldobin:2004:Art-0-pi}.
By creating discontinuities with $\kappa\neq\pi$ ($|\kappa| \leq 2\pi$) one can spontaneously form and study vortices carrying an arbitrary topological charge of $\wp=-\kappa$ or $\wp=-\kappa+2\pi\sgn{\kappa}$ that automatically appear pinned at the $\kappa$-discontinuity. The fractional flux associated with a vortex carrying a topological charge $\wp$ is $\Phi=\Phi_0\,\wp/(2\pi)$. Fractional fluxons can form a variety of ground states~\cite{Goldobin:2KappaGroundStates}, have characteristic eigenfrequencies~\cite{Goldobin:2KappaEigenModes} and get depinned by overcritical bias currents~\cite{Malomed:2004:ALJJ:Ic(Iinj),Goldobin:2004:F-SF}.

{In the present paper we investigate a ``molecule''~\cite{Kienzle:2012:Molecule:Spectroscopy} consisting of two fractional vortices in which the topological charge of the vortices can be tuned electronically by changing $\kappa$. For appropriate parameters of the LJJ we obtain an electronically tunable macroscopic quantum system which {{}}{is suitable for the experimental observation of} coherent quantum oscillations.
}

\subsection{Outline of the article}

Our article is organised as follows. In section~\ref{Sec:Concept} we describe the concept {{}}{of our tunable system} and introduce the model of an LJJ with two discontinuities $\kappa_1$ and $\kappa_2$ separated by a distance $a$. Based on this model we derive in section~\ref{Sec:Solutions} analytical expressions for stationary phases and their corresponding energies. In section~\ref{Sec:Manipulating} we address the question of the preparation of the initial state and show that the energy barrier can be reduced to reach the quantum regime discussed in more detail in section~\ref{Sec:Quantum}. The distinguishability of the two states of the molecule is the topic of section~\ref{Sec:Read-out}. In section~\ref{Sec:Decoherence} we qualitatively discuss the sensitivity of the system to fluctuations and {section~\ref{Sec:Conclusions} summarises our main results.} 
Finally, to keep the paper self-contained, an appendix provides a collection of relations for elliptic functions.

\section{A tunable vortex molecule}
\label{Sec:Concept}

In this section we briefly discuss the concept of a tunable molecule consisting of two fractional Josephson vortices. Moreover, we introduce the model of our system together with the corresponding basic equations.

\subsection{Concept}

Previously~\cite{Goldobin:2005:MQC-2SFs}, we considered molecules consisting of two semifluxons in a $0$-$\pi$-$0$ LJJ situated at the beginning and end of the $\pi$ segment of length $a$. The two states of the molecule were composed of an antiferromagnetically (AFM) arranged pair of a semifluxon and an antisemifluxon. Such a molecule exhibits two degenerate ground states. We found that macroscopic quantum effects can be observed if $a$ is restricted to values~\cite{Goldobin:2005:MQC-2SFs} $a_c<a\lesssim a_c+0.02$, where lengths are measured in units of the Josephson penetration depth $\lambda_J$ and $a_c=\pi/2$ is a bifurcation point. For $a<a_c$ there exists only one static solution with constant phase (flat phase state), while for $a>a_c$ there are two degenerate solutions (AFM states).

It is rather difficult to fabricate the junction with the value of $a$ that lays within the narrow range mentioned above. Therefore, it is highly desirable to make $a$ or $a_c$ tunable. The scaled length $a$ is fixed by design and can vary only slightly due to variations of $\lambda_J$, \eg, with temperature. {For a 0-$\pi$-0 junction, the value of $a_c$ is fixed. For a 0-$\kappa$-$2\kappa$ junction, however, the value of $a_c$ can be tuned electronically on a large range as described in the following.}
\begin{figure}[t]
  \centering
  \includegraphics{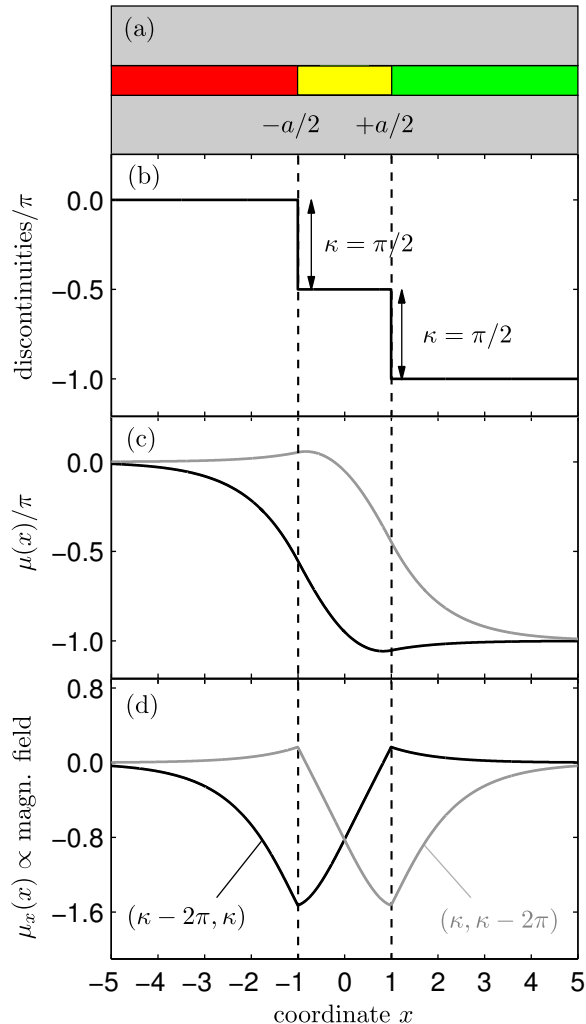}
  \caption{(a) Sketch of a 0-$\kappa$-$2\kappa$ LJJ with $a=2$, where the two discontinuities (b) are located at $x=\pm a/2$. The phases $\mu(x)$ (c) and corresponding magnetic field profiles $\mu_x(x)$ (d) describe two states $(\kappa,\kappa-2\pi)$ and \mbox{$(\kappa-2\pi,\kappa)$}. The coordinate $x$ is normalised to $\lambda_J$.
  }
  \label{Fig:Setup}
\end{figure}

\begin{figure}[t]
  \centering
  \includegraphics{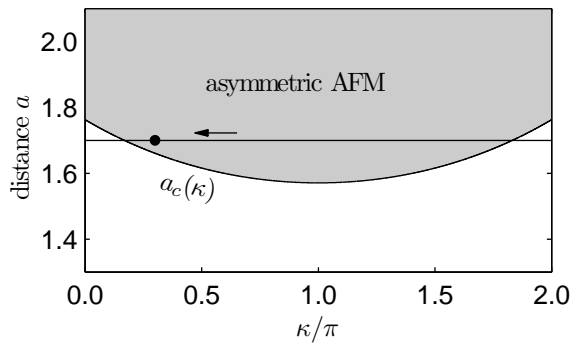}
  \caption{%
    The critical distance $a_c$ for asymmetric AFM molecules as a function of $\kappa$. It separates the flat phase state (white region) from the AFM state (grey region). As an example we consider an LJJ with $a=1.7$ (straight line). By decreasing $\kappa$ we can tune $a_c$ towards $a$, \eg, $\kappa=0.3\pi$ corresponding to $a_c=1.66$ as indicated by the point on the straight line.
  }
  \label{Fig:a_c(kappa)}
\end{figure}

Consider the 0-$\kappa$-$2\kappa$ LJJ of figure~\ref{Fig:Setup} (a) with two discontinuities $\kappa$, see figure~\ref{Fig:Setup} (b). For $\kappa=\pi$ this reduces to the previous case of a 0-$\pi$-$2\pi$ LJJ or, as the Josephson phase is $2\pi$ periodic, to a 0-$\pi$-0 LJJ. For $a>a_c$ the two classically degenerate ground states are molecules consisting of vortices with topological charges $(+\pi,-\pi)$ or $(-\pi,+\pi)$. Now, let us vary $\kappa$ around $\kappa=\pi$. The two corresponding states shown in Figs.~\ref{Fig:Setup} (c) and (d) then are $(\kappa,\kappa-2\pi)$ and $(\kappa-2\pi,\kappa)$~\cite{Goldobin:2KappaGroundStates}. Note that the critical distance~\cite{Goldobin:2KappaGroundStates}
\begin{equation}
  a_c(\kappa)=2\F\left(\frac{\pi}{4},\sqrt{1-\sin\frac{\kappa}{2}}\right),
  \label{Eq:a_c}
\end{equation}
where $\F(\phi,k)$ is the elliptic integral defined in \eqref{Eq:app:ellip:int:first}, is not constant but depends (weakly) on $\kappa$ as shown in figure~\ref{Fig:a_c(kappa)}. We find that the value of $a_c$ may change from $a_c(\pi)=\pi/2\approx1.57$ to $a_c(0)=a_c(2\pi)=2\ln(1+\sqrt{2})\approx1.76$.

Thus, we can fabricate an LJJ with the scaled distance $1.57<a<1.76$ between the discontinuities. Then, we can decrease $\kappa$ starting from $\kappa=\pi$ to bring $a_c(\kappa)$ as close as needed towards $a$, as shown in figure~\ref{Fig:a_c(kappa)}. Note that the required precision for $a$ is less restrictive by one order of magnitude than in the previous proposal~\cite{Goldobin:2005:MQC-2SFs}. In addition we obtain a system which can be tuned into the quantum regime. In the following sections we analyse this system in detail.

\subsection{Model}
\label{Sec:model}

The dynamics of the Josephson phase $\mu(x,t)$ in infinitely long LJJs without bias current and dissipation is described by the sine-Gordon equation~\cite{Goldobin:SF-Shape}
\begin{equation}
  \mu_{xx}(x,t)-\mu_{tt}(x,t)-\sin[\mu(x,t)+\theta(x)]=0
  . \label{Eq:sigo:general}
\end{equation}
We use dimensionless quantities, where lengths are written in units of the Josephson penetration depth $\lambda_J$ and time is written in units of $\omega^{-1}_p$, where $\omega_p$ is the plasma frequency. Energies are measured in units of $E_J\lambda_J$, where $E_J$ is the Josephson energy per length. The subscripts $x$ and $t$ denote partial derivatives with respect to coordinate and time, accordingly.

The function
\begin{equation}
 \theta(x) \equiv -\kappa_1\Heavyside\left(x+\frac{a}{2}\right) - \kappa_2\Heavyside\left(x-\frac{a}{2}\right)
   \label{Eq:theta:def}
\end{equation}
describes two discontinuities $-\kappa_1$ and $-\kappa_2$ located at $x=-a/2$ and $x=+a/2$, where $\Heavyside(x)$ is the Heaviside step function. The values of $\kappa_1$ and $\kappa_2$ can be controlled~\cite{Gaber:2005:NonIdealInj2} by injector currents $I_\mathrm{inj1} \propto \kappa_1$ and $I_\mathrm{inj2} \propto \kappa_2$. The two discontinuities divide the $x$-axis into three parts to which we will refer to as the left, middle and right part of the junction, see figure~\ref{Fig:Setup} (a).

The boundary conditions for $\mu(x,t)$ are given by
\begin{equation}
  \mu_x(-\infty,t)=\mu_x(+\infty,t)=0
  . \label{Eq:mu:bound:orig}
\end{equation}
Although there are discontinuities at $x=\pm a/2$ the phase $\mu(x,t)$ and its derivative $\mu_x(x,t)$ with respect to $x$ have to be continuous at $x=\pm a/2$ at any time $t$.

The sine-Gordon equation (\ref{Eq:sigo:general}) can be derived from the Lagrangian density
\begin{equation}
  \mathcal{L}\equiv {\textstyle\frac{1}{2}}{\mu_t^2}(x,t)-\mathcal{U}
  , \label{Eq:lagrange}
\end{equation}
where
\begin{equation}
  \mathcal{U}\equiv {\textstyle\frac{1}{2}}{\mu_x^2}(x,t)+1-\cos\left[\mu(x,t)+\theta(x)\right]
  \label{Eq:energy:pot:gen}
\end{equation}
is the potential energy density.

To avoid infinite energies the restriction
\begin{equation}
  \cos\left[\mu(\pm\infty,t)+\theta(\pm\infty,t)\right]=1
  \label{Eq:mu:bound:new}
\end{equation}
assures that the potential energy density \eqref{Eq:energy:pot:gen} vanishes at $x=\pm\infty$. This condition restricts the phases at $x=\pm \infty$ to the values
\begin{eqnarray}
  \mu_L&\equiv&\mu(-\infty,t)=2\pi n_L \label{Eq:mu:infty:a} \\
  \mu_R&\equiv&\mu(+\infty,t)= 2\pi n_R + \kappa_1 + \kappa_2 \label{Eq:mu:infty:b}%
  ,
\end{eqnarray}
where $n_L$ and $n_R$ are integer numbers.

\section{Stationary Solutions}\label{Sec:Solutions}

In this section we first derive analytical expressions for the stationary Josephson phase $\mu(x)$ and the corresponding energies. These expressions depend on the values
  \begin{eqnarray}
    \mu_l&\equiv&\mu\left(-{a}/{2}\right)
     \label{Eq:mu:continuity:l}
  \end{eqnarray}
and
  \begin{eqnarray}
    \mu_r&\equiv&\mu\left(+{a}/{2}\right)
    \label{Eq:mu:continuity:r}
  \end{eqnarray}
of $\mu$ at the boundaries (the left and the right) between the regions which have to be determined self-consistently. Furthermore, we analyse the stability of the stationary solutions.

\subsection{Phases}\label{Sec:Phases}

In order to find stationary solutions of (\ref{Eq:sigo:general}) we integrate its static version
\begin{equation}
  \mu_{xx}-\sin(\mu+\theta)=0
  \label{Eq:sigo:general_static}
\end{equation}
in each region of the junction. This procedure leads to a conservation relation
\begin{equation}
  {\textstyle\frac{1}{2}}\mu_x^2+\cos(\mu+\theta)=2k^2-1
  , \label{Eq:conservation_law}
\end{equation}
or
\begin{equation}
 \left(\frac{1}{2k}\mu_x\right)^2=1-\frac{1}{k^2}\sin^2\left(\frac{\mu+\theta}{2}+\frac{\pi}{2}\right)
 , \label{Eq:sol:pre:general}
\end{equation}
where $k$ is a positive real number and has different values in the inner and outer parts of the junction.  

By comparing this equation with~(\ref{Eq:app:a:dn:diff:2}) we see that the phase
\begin{equation}
 \mu(x)=(2n-1)\pi-\theta(x)\pm2\am\left[k(x+c),k^{-1}\right]
 \label{Eq:sol:general}
\end{equation}
solves~(\ref{Eq:sigo:general_static}) in every region. Here $n$ is an integer number and $c$ is an arbitrary constant. Furthermore we have introduced the Jacobi amplitude function $\am(u,k)$ defined by~(\ref{Eq:app:am_F}) and (\ref{Eq:app:a:1:am}).

From the boundary conditions (\ref{Eq:mu:bound:orig}) and (\ref{Eq:mu:bound:new}) in~(\ref{Eq:conservation_law}) we find
\begin{equation}
 k=1
  \label{Eq:b=1}
\end{equation}
in the left and right part of the junction.

By matching~(\ref{Eq:conservation_law}) for the outer part ($k=1$) and the inner part ($k\neq 1$) of the junction at $x=\pm a/2$ we obtain
\begin{eqnarray}
  k^2&=&1+\sin\frac{\kappa_1}{2}\sin\left(\mu_l-\frac{\kappa_1}{2}\right)
  \nonumber\\
  &=&1{\color{black}-\sin\frac{\kappa_2}{2}\sin\left(\mu_r-\kappa_1-\frac{\kappa_2}{2}\right)}
  . \label{Eq:k}
\end{eqnarray}
Therefore, $k$ can vary between the values 0 and $\sqrt{2}$ in the middle of the junction. Since the behaviour of the Jacobi amplitude function depends on the value of $k$ as discussed in the appendix, we consider three different cases: (a) outer regions with $k=1$, (b) inner region with $k > 1$ and (c) inner region with $k < 1$.

\paragraph{Outer regions with $k=1$.} With the help of \eqref{Eq:app:k=1:am} we obtain from~\eqref{Eq:sol:general} the expression
\begin{eqnarray}
  \mu(x)&=&\mu_{L,R}+4\arctan\Big(\ee^{a/2-|x|}
  \tan\frac{\mu_{l,r}-\mu_{L,R}}{4}\Big)
 \label{Eq:mu:outer:main}
\end{eqnarray}
for $k \to 1$, where $\mu_{l}$ and $\mu_{r}$ are restricted to
\begin{eqnarray}
 |\mu_{l,r}-\mu_{L,R}|<2\pi
 . \label{Eq:mu:restriction}
\end{eqnarray}

Since we have to match these two solutions and their derivatives to their counterparts in the inner region we define the signs
\begin{eqnarray}
  \sigma_{l}&\equiv&\mathrm{sign}[\mu_x(-a/2)]=\mathrm{sign}(\mu_l-\mu_{L}),\\
  \sigma_{r}&\equiv&\mathrm{sign}[\mu_x(+a/2)]=\mathrm{sign}(\mu_R-\mu_{r})   \label{Eq:sigma:def}%
\end{eqnarray}
of $\mu_x(x)$.

\paragraph{Inner region with $k> 1$.} With the help of~\eqref{Eq:app:am_F} it can easily be verified  that
\begin{eqnarray}
 \mu(x)&=&\kappa_1-\pi+\nonumber\\
  &&2\,\sigma\am\left[k\left(x+{a}/{2}\right)+\sigma \F(\psi_{l},k^{-1}),k^{-1}\right]
  \label{Eq:mu:middle:mon}
\end{eqnarray}
matches the solution for the left part of the junction, that is $\mu(-a/2)=\mu_l$ and $\sigma=\sigma_l$. Here we have used the abbreviation
\begin{eqnarray}
   \psi_{l,r}&\equiv&{\textstyle\frac{1}{2}}\left({\mu_{l,r}+\pi-\kappa_1}{}\right)
  .
\end{eqnarray}

To match the solution for the right part of the junction, that is $\mu(a/2)=\mu_r$ and $\sigma=\sigma_r$, we have the additional constraint
\begin{equation}
  a={\sigma}k^{-1}\left[\F(\psi_r,k^{-1})-\F(\psi_l,k^{-1})\right]
  . \label{Eq:a:mon}
\end{equation}

\paragraph{Inner region with $k<1$.} Similar to the case $k>1$ we find
\begin{eqnarray}
 \mu(x)&=&{(2n-1)\pi}+\kappa_1+\nonumber\\
  &&2\sigma_l\am [k(x+{a}/{2})+\sigma_l\,k\F(\varphi_{l},k),k^{-1}]
  , \label{Eq:mu:middle:peri:final}
\end{eqnarray}
where we have used
\begin{eqnarray}
  \varphi_{l,r}&\equiv& \arcsin\left(k^{-1}\sin\bar\psi_{l,r}\right)
\end{eqnarray}
and
\begin{eqnarray}
    \bar\psi_{l,r}&+&n\pi \equiv\psi_{l,r},\quad|\bar \psi_{l,r}|\leq\pi/2
  . \label{Eq:n:def:main}
\end{eqnarray}
Note that there is only one integer number $n$. Therefore, the values of $\psi_{l,r}$ and $\mu_{l,r}$ are restricted to
$|\psi_r - \psi_l| \le \pi$ and $|\mu_r - \mu_l| \le 2\pi$.

The additional constraint to match the solution for the right part reads
\begin{eqnarray}
  a&=&\left[\sigma_r\F(\varphi_{r},k)-\sigma_l\F(\varphi_{l},k)\right] \nonumber\\
  &&+(1-\sigma_l\sigma_r)\K(k)+4\nu\K(k)
  . \label{Eq:a:middle:peri:final}
\end{eqnarray}
Due to the properties of elliptic integrals, for a given value of $a$ the integer number $\nu$ is limited to the values
\begin{equation}
  0\leq \nu < \frac{a}{2\pi}+\frac{1}{2}
  . \label{Eq:m:def:main:app}
\end{equation}

We now have expressed the stationary solutions of the sine-Gordon equation \eqref{Eq:sigo:general} in terms of $\mu_L$, $\mu_R$, $\mu_l$ and $\mu_r$. The possible values of $\mu_L$ and $\mu_R$ are restricted by~\eqref{Eq:mu:infty:a} and \eqref{Eq:mu:infty:b} and depend on the topological charge of the system. The values of $\mu_l$ and $\mu_r$ have to be determined from~\eqref{Eq:k}, \eqref{Eq:a:mon} and \eqref{Eq:a:middle:peri:final}. Examples will be presented in section~\ref{Sec:Tuning}.

\subsection{Energies}\label{Sec:energies}

In this section we use the results of the previous section to calculate the energy of the stationary solutions.

Integrating~\eqref{Eq:energy:pot:gen} and using \eqref{Eq:conservation_law} we obtain the energy
\begin{equation}
 U_l=\int_{-\infty}^{-a/2} \mu_x^2\ \dd x= \int_{\mu_L}^{\mu_l} \mu_x\ \dd \mu
  \label{Eq:energy:general:outer}
\end{equation}
of the solution in the left part of the junction. From~\eqref{Eq:mu:outer:main}  we find
\begin{equation}
 \mu_x=2 \sin \left[ {\textstyle \frac{1}{2}}\left(\mu(x)-\mu_L{}\right) \right]
\end{equation}
and arrive at
\begin{eqnarray}
  U_{l}&=&8\sin^2 \left[\textstyle \frac{1}{4} \left( {\mu_{l}-\mu_{L}}{} \right) \right]
  .  \label{Eq:energy:outer:l}
\end{eqnarray}
Similarly, we obtain the expression
\begin{eqnarray}
  U_{r}&=&8\sin^2 \left[\textstyle \frac{1}{4} \left( {\mu_{r}-\mu_{R}}{} \right) \right]
    \label{Eq:energy:outer:r}
\end{eqnarray}
for the right part of the junction.

For the middle part of the junction we again combine~\eqref{Eq:energy:pot:gen} and \eqref{Eq:conservation_law} and obtain the energy
\begin{equation}
 U_m=\int_{-a/2}^{a/2} \left[ \mu_x^2+2(1-k^{2}) \right]\,\dd x
 \label{Eq:energy:general:inner}
\end{equation}
of the phase in this region of the junction. From~\eqref{Eq:sol:general} we find
\begin{equation}
 \mu_x^2=4k^2 \dn^2 \left[k(x+c),k^{-1}\right]
  \label{Eq:energy:mux2}
\end{equation}
and
\begin{equation}
 \pm \sn \left[k(x+c),k^{-1}\right] = (-1)^n \cos \left[\textstyle \frac{1}{2} \left( {\mu(x)-\kappa_1}{} \right) \right]
  . \label{Eq:energy:sn:pm}
\end{equation}

Equations \eqref{Eq:energy:mux2} and \eqref{Eq:app:epsilon:1} allow us to evaluate the integral. Finally, we obtain with the help of~\eqref{Eq:app:epsilon:add} and \eqref{Eq:energy:sn:pm} the energy
\begin{eqnarray}
 U_m&=&2a(1-k^{2})+4k\mathcal{E}\left(ka,k^{-1}\right) \nonumber\\
  &&- {4} k^{-1} \cos\frac{\mu_l-\kappa_1}{2} \cos\frac{\mu_r-\kappa_1}{2}\sn(ka,k^{-1})
  , \label{Eq:energy:middle:general}
\end{eqnarray}
where $\mathcal{E}$ is the Jacobi epsilon function defined by~\eqref{Eq:app:epsilon:1}.
The total energy
\begin{equation}
  U=U_l+U_m+U_r
   \label{Eq:energy:final}
\end{equation}
of a stationary solution $\mu(x)$ is the sum of the energies of these three domains.

\subsection{Stability analysis}\label{Sec:eigenmodes}

To study the stability of a stationary solution $\mu(x)$ we insert the ansatz
\begin{equation}
  \mu(x,t)=\mu_{}(x)+\psi(x)e^{-\ii\omega t}
\end{equation}
into the sine-Gordon equation (\ref{Eq:sigo:general}) and linearise it around the stationary solution $\mu(x)$. Since $\mu_{}(x)$ solves the stationary sine-Gordon equation (\ref{Eq:sigo:general_static}), we obtain the differential equation
\begin{equation}
  -\psi''(x)+\cos\left[\mu_{}(x)+\theta(x)\right]\psi(x)=\omega^2\psi(x)
   \label{Eq:Schroedinger}
\end{equation}
for the eigenmodes $\psi(x)$ with eigenvalues $\omega^2$, which, after substituting~\eqref{Eq:sol:general}, takes the form
\begin{equation}
  \psi''(x)+[\,1+\omega^2-2\sn^2(k(x+c),k^{-1})]\psi(x)=0
  . \label{Eq:Lame}
\end{equation}
Note that~\eqref{Eq:Lame} is the Lam\'{e} equation~\cite{nist2011}. The boundary conditions and the matching conditions at $x=\pm a/2$ for $\psi(x)$ follow from the boundary and matching conditions for $\mu(x)$ discussed in section~\ref{Sec:model}.

As long as the lowest eigenvalue $\omega_0^2$ of the Lam\'{e} equation (\ref{Eq:Lame}) is positive, the stationary solution $\mu_{}(x)$ is stable. When it becomes negative, the stationary solution $\mu_{}(x)$ is unstable. In section~\ref{Sec:Tuning} we solve~\eqref{Eq:Lame} numerically.

\section{Tailoring molecule states}
\label{Sec:Manipulating}

Our ultimate goal is {{}}{to obtain a system which shows experimentally observable coherent quantum oscillations between the two classical degenerate ground states of an AFM molecule.} For this purpose we first bring the system into one of its degenerate ground states and choose $\kappa$ such that the barrier between the two states is high. By tuning $\kappa$ we then reduce this energy barrier to reach the quantum regime and calculate the corresponding scaled energy splitting $\delta \varepsilon_{01}$.

\subsection{Preparing an initial AFM state}
\label{Sec:Initializing}

In a previous publication~\cite{Goldobin:2005:MQC-2SFs} it was shown that a configuration with the discontinuities $\kappa_1=\pi$ and $\kappa_2=-\pi$ ($0$-$\pi$-$0$ LJJ) provides an effective double-well potential with degenerate ground states where quantum tunnelling can be observed.

An intuitive way to arrive at this state from the \mbox{$\kappa_1=\kappa_2=0$} state is to use
\begin{equation}
  \kappa_1=\kappa_i, \quad \kappa_2=-\kappa_i
   \label{Eq:disc:init:2}
\end{equation}
and sweeping $\kappa_i$ from $0$ to $\pi$. Then, we want to deal with a
molecule consisting of a direct and a complementary vortex. Therefore, we have to change the discontinuities appropriately, \eg,
\begin{equation}
  \kappa_1=\pi+\kappa_t, \quad \kappa_2=-\pi+\kappa_t
   , \label{Eq:disc:tune:2}
\end{equation}
where the (de)tuning $\kappa_t$ starts at $0$ and changes in the range between $-\pi$ and $+\pi$.

In this approach one can employ two injector current sources: one creating $\kappa_i$ wired to produce two discontinuities of opposite signs like in~\eqref{Eq:disc:init:2} and another one creating $\kappa_t$ and wired to produce two discontinuities of the same sign like in~\eqref{Eq:disc:tune:2}. The advantages of this approach are: instead of using the first current source, one could replace the junction by a conventional 0-$\pi$-0 LJJ and then only apply injector currents for the tuning $\kappa_t$ according to~\eqref{Eq:disc:tune:2};~(ii) this technique works also in an annular JJ;~(iii) after $\kappa_i$ has reached the value $\pi$ we know that we have prepared the $(+\pi,-\pi)$ state rather than the $(-\pi,+\pi)$ state. The disadvantage of this approach is that one always injects a current $\propto\kappa=\pi$ or more into the system, which brings additional noise.

An alternative way is to use only one injector current source, which is coupled to injectors as
\begin{equation}
    \kappa_1=\kappa_2=\kappa
 .\label{Eq:disc:init:3}
\end{equation}
Starting from the $\kappa=0$ state and no vortices in the JJ, one slowly increases the value of $\kappa$ and finds the symmetric ferromagnetic vortex configurations $(\kappa,\kappa)$. When $\kappa$ is increased above $\kappa_c^{\upharpoonleft\upharpoonleft}(a)$~\cite{Goldobin:2KappaGroundStates}, where $\pi<\kappa_c^{\upharpoonleft\upharpoonleft}(a)<2\pi$, the molecule emits one fluxon and reconfigures into one of the degenerate asymmetric antiferromagnetic states $(\kappa,\kappa-2\pi)$ or $(\kappa-2\pi,\kappa)$. To reach the $(+\pi,-\pi)$ or $(-\pi,+\pi)$ state we decrease $\kappa$ to $\pi$. Then we further decrease $\kappa$ to reach the quantum limit. This technique has two advantages:~(i) one needs only one injector current source;~(ii) one can operate the quantum system at small $\kappa$ values, which brings less noise. This technique, however, has two disadvantages:~(i) it does not work in annular LJJs, where the emitted fluxon cannot leave easily\footnote{One may also think about a 
fluxon trap --- that is a
potential well created by a magnetic field or junction width modulation~\cite{Goldobin:RatchetT:2001}, which will keep the emitted fluxon far from the molecule. Another possibility is to implement a fluxon absorbing injector pair far from the molecule.};~(ii) we do not know if we prepared the $(\kappa,\kappa-2\pi)$ or $(\kappa-2\pi,\kappa)$ state.

For our further calculations we choose the second approach where $\kappa$, and therefore, the additional noise in the system is small in the quantum limit.

\subsection{Tuning the barrier height}\label{Sec:Tuning}

We consider the particular injector-tuning technique described by~\eqref{Eq:disc:init:3} for characterising the states, its energies and stabilities during this process.

\subsubsection{Determination of phase parameters}

After the preparation procedure discussed in the previous section the two discontinuities have the value $\kappa=\pi$ and the AFM state has the topological charge \mbox{$\mu_R-\mu_L=0$}. For simplicity we choose $n_L=0$ and $n_R=-1$ in~\eqref{Eq:mu:infty:a} and \eqref{Eq:mu:infty:b}.

When we tune $\kappa$ no (anti)fluxon is emitted or absorbed. Therefore, $n_L$ and $n_R$ do not change and the topological charge
\begin{equation}
  \mu_R-\mu_L=2\kappa-2\pi
    \label{Eq:muR:2kappa}
\end{equation}
is induced in the system when the discontinuities have the value $\kappa$.

In the stationary solutions presented in section~\ref{Sec:Solutions} we still have to find $\mu_l$ and $\mu_r$. In our examples we restrict ourselves to the values of $a$ covering the grey region in figure~\ref{Fig:a_c(kappa)} near the boundary $a_c(\kappa)$, that is, $1.5\le a \le 2$ and $0\le \kappa \le 2\pi$. It turns out that for these parameters there are only solutions for $k<1$. For this limitation,~\eqref{Eq:k} restricts the values of $\mu_l$ and $\mu_r$ to the intervals
\begin{eqnarray}
  \frac{\kappa}{2}-\pi<&\mu_{l}&+2\pi n_1<\frac{\kappa}{2}
  \label{Eq:mu:restr:tune:high:a}%
\end{eqnarray}
and
\begin{eqnarray}
  \frac{3\kappa}{2}<&\mu_{r}&+2\pi n_2<\pi+\frac{3\kappa}{2}
  . \label{Eq:mu:restr:tune:high:b}%
\end{eqnarray}
{\color{black}Here we use instead of~\eqref{Eq:mu:restriction} the more restrictive condition $|\mu_{l,r}-\mu_{L,R}|<\pi$ which requires $n_1=0$ and $n_2=1$. This condition is reasonable because even in the case when we have a whole integer fluxon at one of the discontinuities the phase changes at most by $\pi$ when one goes from infinity up to the fluxon centre.}

By using the constraints~\eqref{Eq:mu:restr:tune:high:a} and \eqref{Eq:mu:restr:tune:high:b} together with~\eqref{Eq:k} we express $\mu_r$ in terms of $\mu_l$ and obtain
\begin{equation}
  \mu_r=-\mu_l+2\kappa-2\pi
  \label{Eq:case:tune:inst}
\end{equation}
and
\begin{equation}
  \mu_r=\mu_l+\kappa-\pi
    \label{Eq:case:tune:stab}
\end{equation}
as the two only possible solutions.

\begin{figure}[b]
  \centering
  \mbox{\includegraphics{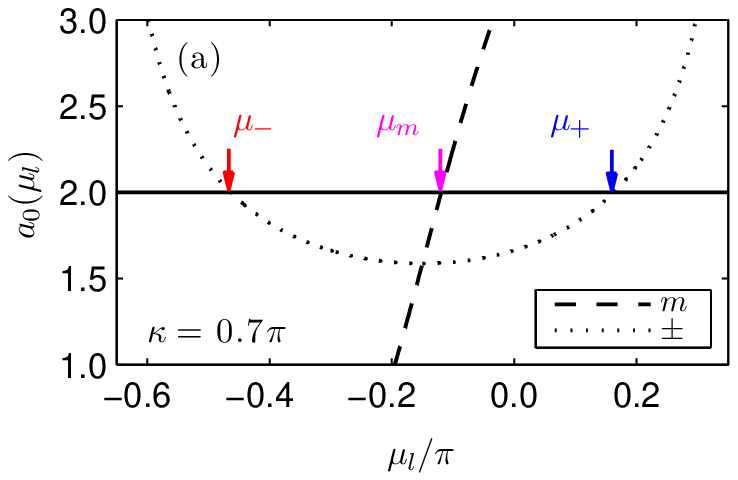}\includegraphics{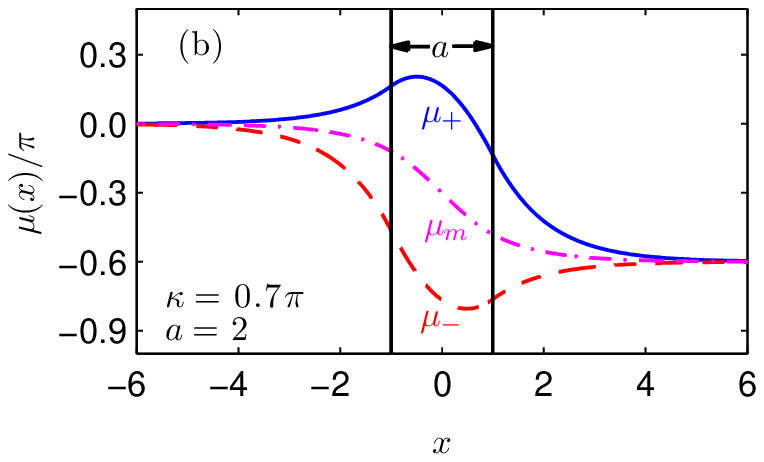}}
  \caption{(a) Possible values of $\mu_l$ for $a=2$ and \mbox{$\kappa=0.7\pi$}. When we insert~\eqref{Eq:case:tune:inst} and \eqref{Eq:case:tune:stab} into~(\ref{Eq:a:middle:peri:final}) we obtain for $\nu=0$ two functions $a_0(\mu_l)$ depicted by a dotted line and by a dashed line, respectively. For a given value $a=2$ three values for $\mu_l$ are possible: $\mu_l=\mu_m$ for~\eqref{Eq:case:tune:inst} and $\mu_l=\mu_\pm$ for~\eqref{Eq:case:tune:stab}. (b) The phases $\mu_m(x)$, $\mu_+(x)$ and $\mu_-(x)$ corresponding to the values $\mu_l=\mu_m$, $\mu_l=\mu_+$ and $\mu_l=\mu_-$ (arrows in (a)) for $a=2$ and $\kappa=0.7\pi$.
  }
  \label{Fig:a:tune}
  \label{Fig:phases:tune}
\end{figure}

Next we substitute these two expressions into~\eqref{Eq:a:middle:peri:final} and solve the resulting equations numerically for $\mu_l$ for a given distance $a$. We denote the solution corresponding to~\eqref{Eq:case:tune:inst} by an index $m$, and the two solutions corresponding to~\eqref{Eq:case:tune:stab} by an index $+$ and $-$. Figure~\ref{Fig:a:tune} (a) illustrates this process for $a=2$.

When we insert the three values of $\mu_m$ and $\mu_\pm$ into~\eqref{Eq:mu:outer:main} and \eqref{Eq:mu:middle:peri:final} we find the phases $\mu_m(x)$ and $\mu_\pm(x)$ depicted in figure~\ref{Fig:phases:tune} (b).
These solutions satisfy the symmetry relations
\begin{equation}
 \mu_m(-x) = -\mu_m(x) +2\kappa -2\pi
  \label{Eq:inst:sym}
\end{equation}
and
\begin{equation}
 \mu_{\pm}(-x) = -\mu_{\mp}(x) +2\kappa -2\pi
  . \label{Eq:stab:sym}
\end{equation}%
The solution $\mu_+(x)$ describes a molecule consisting of a $\kappa$ vortex at $x=-a/2$ and the complementary $\kappa-2\pi$ vortex at $x=+a/2$. The solution $\mu_-(x)$ is a complementary molecule $(\kappa-2\pi,\kappa)$, which has the same energy. These two solutions are stable whereas the solution $\mu_m(x)$ is unstable, see section~\ref{Sec:En&EigenFreq}.

\subsubsection{Energies and stability analysis}
\label{Sec:En&EigenFreq}

Depending on the parameters of the LJJ there is either a single stable stationary solution $\mu_m(x)$, or two stable solutions $\mu_\pm(x)$ and one unstable solution $\mu_m(x)$. The two stable solutions are separated by an energy barrier which is governed by the unstable solution. To reach the quantum limit we want to reduce this energy barrier by tuning $\kappa$. Therefore, we investigate how the energies and stabilities of the stationary solutions depend on $\kappa$.

For our examples we use the following JJ parameters:
\begin{equation}
  \omega_p=\sqrt{\frac{2\pi j_c}{\Phi_0 C}},\quad E_J=\frac{j_c w \Phi_0}{2\pi}
  ,\quad
  \lambda_J=\sqrt{\frac{\Phi_0}{2\pi \mu_0 2 \lambda_L j_c}}
  , \label{Eq:units}
\end{equation}
where $\lambda_L$ is the London penetration depth, $j_c$ is the critical current density of the JJ, $C$ is the capacitance of the JJ per unit of area, $w$ is the JJ's width, and $\Phi_0\approx2.07\times10^{-15}\units{Wb}$. Typical parameters are $\lambda_L=90\units{nm}$, $w = 1\units{\mu m}$, and $C= 4.1\units{\mu F/cm^2}$ with $j_c = 100\units{A/cm^2}$. This corresponds to $\lambda_J=38\units{\mu m}$, $\omega_p = 2\pi \times 42.8 \units{GHz}$, and \mbox{$E_J \lambda_J=78.4\units{meV}$}.

Without losing generality we restrict ourselves to \mbox{$0<\kappa<\pi$}. The results for the range $\pi<\kappa<2\pi$ are obtained by substituting $\kappa\to2\pi-\kappa$.

\begin{figure}[b]
  \centering
  \mbox{\includegraphics{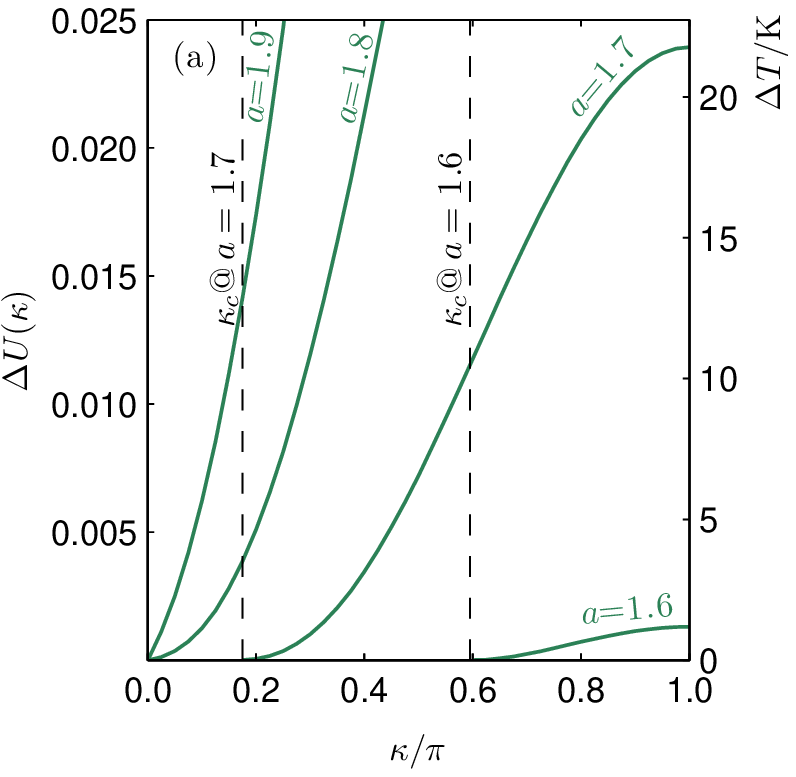}\includegraphics{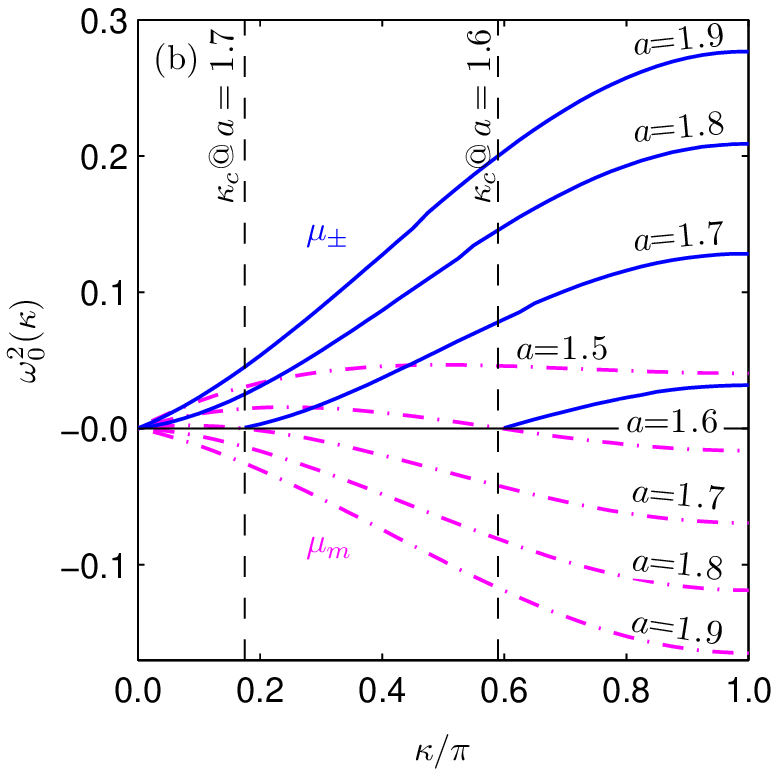}}
  \caption{(a) Energy difference $\Delta U(\kappa)$~\eqref{Eq:energy:diff} for different values of $a$. By tuning $\kappa$, $\Delta U(\kappa)$ can be made as small as needed. The right vertical axis shows the corresponding temperature $\Delta T=\Delta U\cdot E_J\lambda_J/k_B$ for typical experimental parameters mentioned after~\eqref{Eq:units}. (b) Lowest eigenvalue $\omega_0^2(\kappa)$ for the stationary solutions $\mu_m(x)$ and $\mu_{\pm}(x)$. When $\omega_0^2(\kappa)$ is positive, the stationary solution is stable; when $\omega_0^2(\kappa)$ is negative, it is unstable.
  }
  \label{Fig:energy:tune}
  \label{Fig:frEq:tune}
\end{figure}

In order to calculate the energies of the solutions we insert the values for $\mu_l$ obtained in the previous subsection into~(\ref{Eq:energy:final}) and obtain the energies $U_m(\kappa)$ of the state $\mu_m(x)$ and $U_+(\kappa)=U_-(\kappa)$ of the states $\mu_\pm(x)$. In figure~\ref{Fig:energy:tune} (a) we depict the corresponding energy difference                                                                                                                                                                                                                                                                                                                                                          %
\begin{equation}
  \Delta U(\kappa)\equiv U_m(\kappa)-U_{\pm}(\kappa)
  \label{Eq:energy:diff}
\end{equation}
for different values of $a$. This figure shows that the energy barrier can be tuned down to zero by decreasing the discontinuity $\kappa$ from $\pi$ to 0.

The stability of the solutions $\mu_m(x)$ and $\mu_\pm(x)$ is determined by the eigenvalues of~\eqref{Eq:Lame}. These eigenvalues are calculated numerically. We have used a junction of length $l=20$ to emulate an infinitely long JJ. The results are depicted in figure~\ref{Fig:frEq:tune} (b). Positive eigenvalues characterise stable solutions, whereas negative eigenvalues characterise unstable solutions.

In the lower part of figure~\ref{Fig:a_c(kappa)}, that is for $a<a_c(\pi)\approx1.57$, there is only one stationary solution $\mu_m(x)$, so that $\Delta U(\kappa)\equiv 0$ in figure~\ref{Fig:energy:tune} (a). Figure \ref{Fig:frEq:tune} (b) shows that the corresponding eigenvalue $\omega_0^2$ is positive, therefore this solution is stable.

For $a_c(\pi)<a<a_c(0)\approx 1.76$ we have to distinguish between $\kappa>\kappa_c(a)$ and $\kappa<\kappa_c(a)$, where the function $\kappa_c(a)$ is the inverse of $a_c(\kappa)$ defined by~\eqref{Eq:a_c}. For $\kappa>\kappa_c(a)$ there are three stationary solutions $\mu_m(x)$ and $\mu_\pm(x)$. As shown in figure~\ref{Fig:energy:tune} (a) the energy barrier $\Delta U(\kappa)$ becomes lower when we decrease $\kappa$ and vanishes at $\kappa=\kappa_c(a)$. Our numerics shows that $\Delta U(\kappa)\propto (\kappa-\kappa_c)^{2}$ for $\kappa\to \kappa_c$. From the eigenvalues $\omega_0^2(\kappa)$ displayed in figure~\ref{Fig:frEq:tune} (b) corresponding to these stationary solutions we conclude that $\mu_m(x)$ is unstable whereas $\mu_\pm(x)$ are stable.

At $\kappa=\kappa_c(a)$ all three solutions have the same energy and $\Delta U(\kappa)$ vanishes. The eigenvalues $\omega_0^2(\kappa)$ join at the bifurcation point $\kappa=\kappa_c(a)$ and vanish, see figure~\ref{Fig:frEq:tune} (b). For $\kappa<\kappa_c(a)$ the two stable solutions $\mu_\pm(x)$ disappear, while the unstable solution $\mu_m(x)$ becomes stable, see figure~\ref{Fig:frEq:tune} (b).

Finally, for $a>a_c(0)$ there are always three stationary solutions $\mu_m(x)$ and $\mu_\pm(x)$. According to figure~\ref{Fig:frEq:tune} (b), $\mu_m(x)$ is unstable and $\mu_\pm(x)$ are stable. All three solutions reach the same energy at $\kappa\to 0$ and the energy barrier vanishes as $\Delta U(\kappa)\propto\kappa$, see figure~\ref{Fig:energy:tune} (a). The eigenvalues $\omega_0^2(\kappa)$ corresponding to the three solutions vanish too at $\kappa\to0$.

In this limiting case, all three solutions are just single fluxons. The two stable solutions $\mu_+(x)$ and $\mu_-(x)$ are fluxons weakly pinned at $-a/2$ and $+a/2$, respectively. The unstable solution $\mu_m(x)$ is a fluxon at $x=0$. It is rather interesting that $\Delta U(\kappa)$ vanishes for $\kappa\to0$ even for $a>a_c(0)$. Therefore, one can always make the barrier as small as needed for \emph{arbitrary} $a>a_c(\pi)=\pi/2$ and is \emph{not} limited by the interval of $1.57\lesssim a\lesssim 1.76$!

This limit $\kappa \to 0$ is similar to the heart-shaped {qubit} in which a fluxon is trapped in a double-well potential created by a non-uniform magnetic field~\cite{Kemp:2002:VortexQubit}. In this situation, residual pinning in the system can play a major role and make the parasitic pinning potential larger than the one we construct here.

\section{Quantum regime}
\label{Sec:Quantum}

We expect to observe coherent quantum oscillations between the two degenerate states corresponding to the classical states $\mu_+(x)$ and $\mu_-(x)$ in the region where the energy barrier in figure~\ref{Fig:energy:tune} (a) becomes sufficiently small, \ie, the coupling between these two states is sufficiently large. In order to quantify this effect we map the dynamics of the system to the dynamics of a single particle in a double-well potential and calculate the energy splitting $\delta\varepsilon_{01}$ separating the two lowest energy levels. Because the oscillation frequency between the two states $\mu_+(x)$ and $\mu_-(x)$ is given by $\delta\varepsilon_{01}/\hbar$ we are then able to determine the values of the discontinuity $\kappa \propto I_\mathrm{inj}$ for a given value of $a$ where the quantum oscillations become observable.

\subsection{Single-mode approximation}
\label{Sec:Onemode}

In order to map the dynamics of the complete system to the dynamics of a single particle in a double-well potential we express the phase
\begin{equation}
 \mu(x,t)=\mu_m(x)+\sum_{n=0}^\infty q_n(t)\psi_n(x)
   \label{Eq:expansion}
\end{equation}
in terms of the stationary solution $\mu_m(x)$ and the corresponding eigenmodes $\psi_n(x)$ defined by~(\ref{Eq:Lame}). By inserting this expansion into the Lagrangian density \eqref{Eq:lagrange} and integrating over $x$ we obtain a Lagrangian for the mode amplitudes $q_n(t)$ which describes the motion of a fictitious particle in many dimensions.

If the lowest eigenvalue $\omega_0^2$ of~\eqref{Eq:Lame} is sufficiently separated from the next higher eigenvalue we can expect that for low energies the particle only moves along the ``$q_0$ direction''. Motivated by this simple picture, we omit the higher modes in~\eqref{Eq:expansion} and use the approximation
\begin{equation}
 \mu(x,t)\approx\mu_m(x)+q_0(t)\psi_0(x)
  ,
\end{equation}
where $\psi_0(x)$ is the eigenfunction corresponding to the lowest eigenvalue $\omega_0^2$. We insert this equation into our Lagrangian density \eqref{Eq:lagrange} and take into account only terms up to the fourth order in $q_0$ and obtain the Lagrangian
\begin{equation}
  L={\textstyle \frac{1}{2}} \dot{q_0}^2-U(q_0)
  , \label{Eq:lagrange:q_0}
\end{equation}
where
\begin{equation}
 U(q_0)\equiv {\textstyle \frac{1}{2}}{\omega_0^2}{}q_0^2+{\textstyle\frac{1}{24}}Kq_0^4
  \label{Eq:quartic}
\end{equation}
is the potential energy. Here the positive parameter $K$ is defined by
\begin{eqnarray}
  K&\equiv&-\int_{-\infty}^{+\infty}\dd x\,\cos[\mu_0(x)+\theta(x)]\,\psi_0^4(x)
\end{eqnarray}
and  $\psi_0$ is normalised according to $\int\psi_0^2\,\dd x=1$. Due to the symmetry relation \eqref{Eq:inst:sym} for $\mu_m(x)$ there is no third-order term in~\eqref{Eq:quartic}.

Note that for $\omega_0^2<0$, where $\mu_m(x)$ is unstable, $U(q_0)$ describes a double-well potential with a maximum at $q_0=0$ and two minima at $q_0=\pm\sqrt{6/K}|\omega_0|$. For $\omega_0^2>0$ it only has one minimum at $q_0=0$. In the first case the oscillation frequencies around the minima are $\omega_\pm=\sqrt{2}\,|\omega_0|$. In the second case $\omega_0$ is the oscillation frequency around the minimum.

\subsection{Energy splitting}
\label{Sec:Splitting}

For the Lagrangian \eqref{Eq:lagrange:q_0} we can derive the stationary Schr\"{o}dinger equation
\begin{equation}
 \left(\frac{-\hbar_\mathrm{eff}^2}{2}\frac{\partial^2}{\partial q_0^2} +U(q_0)\right)u_j(q_0)=\varepsilon_j u_j(q_0)
  , \label{Eq:Schroedinger:quartic}
\end{equation}
where
\begin{equation}
 \hbar_\mathrm{eff}\equiv \frac{\hbar \omega_p}{E_J\lambda_J}
\end{equation}
is the dimensionless Planck constant and the energy eigenvalues $\varepsilon_j$ are given in units of $E_J\lambda_J$.

In order to calculate the energy splitting we solve~\eqref{Eq:Schroedinger:quartic} numerically for the potential $U(q_0)$ given by~(\ref{Eq:quartic}). Additionally, we compare our numerical results for $\delta\varepsilon_{01} \equiv \varepsilon_1-\varepsilon_0$ with the scaled energy splitting
\begin{equation}
 \Delta=8\hbar_\mathrm{eff}\sqrt{\frac{2\Delta U}{\pi\hbar_\mathrm{eff}\omega_\pm}}\exp\left(-\frac{16\Delta U}{3\hbar_\mathrm{eff}\omega_\pm}\right)
  \label{Eq:semiclassical}
\end{equation}
in the semiclassical limit~\cite{weiss:1999}.

\begin{figure}[b]
  \centering
  \includegraphics[width=15.7cm]{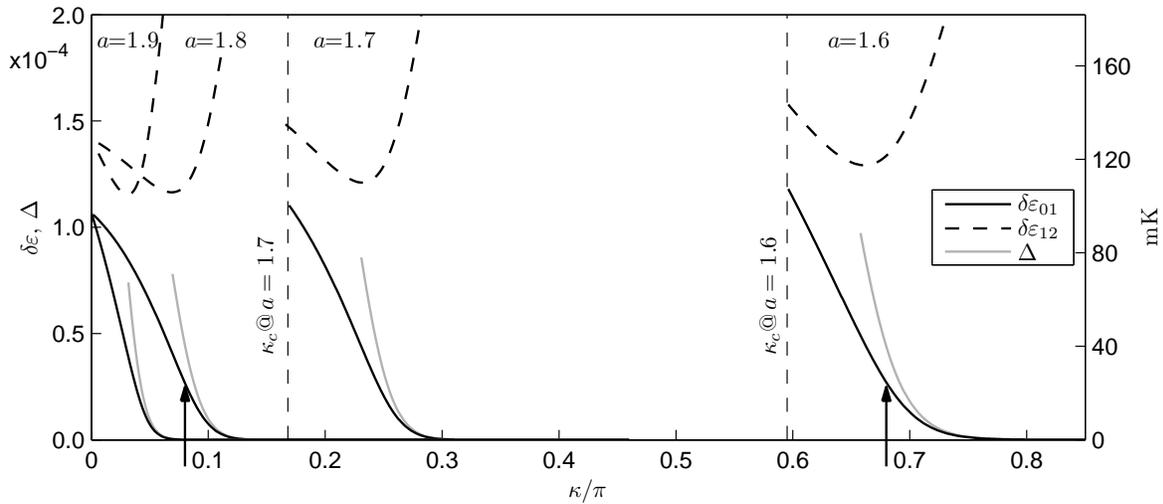}
  \caption{%
    The energy differences $\delta\varepsilon_{01}\equiv \varepsilon_1-\varepsilon_0$ (solid lines) and $\delta\varepsilon_{12}\equiv \varepsilon_2-\varepsilon_1$ (dashed lines) calculated from~\eqref{Eq:Schroedinger:quartic} and the semiclassical expression $\Delta$ (grey lines),~(\ref{Eq:semiclassical}), as a function of $\kappa$ for different values of $a$. Similar as in figure~\ref{Fig:energy:tune} (a) the right vertical axis shows the corresponding temperature for typical parameters mentioned after~\eqref{Eq:units}. For the parameters indicated by two arrows more details are shown in figure~\ref{Fig:potential}.
}
  \label{Fig:quantum}
\end{figure}

Figure \ref{Fig:quantum} shows our numerical results for $\delta\varepsilon_{01}$ and the semi-classical expression $\Delta$ for different values of $a$. We note that the semi-classical expression $\Delta$ is a good approximation for $\delta \varepsilon_{01}$ as long as $\delta \varepsilon_{01}$ is small. To have a good quantum-mechanical two-level system the two lowest eigenvalues $\varepsilon_0$ and $\varepsilon_1$ have to be well separated from the higher eigenvalues. Therefore, we additionally compare $\delta\varepsilon_{12} \equiv \varepsilon_2-\varepsilon_1$ to $\delta \varepsilon_{01}$.

To establish a two-level system at a temperature $T$ three conditions have to be fulfilled:~(i) $\Delta U\gg k_B T$ to suppress thermal hopping between the two classical ground states;~(ii) $\delta \varepsilon_{01}\gg k_BT$ to observe coherent oscillations;~(iii) $\delta \varepsilon_{12}\gg\delta \varepsilon_{01}$ to have approximately a two-level system. For large values of $\Delta U$, the energy splitting $\delta \varepsilon_{01}$ becomes small. Therefore, we have to find parameters where $\Delta U$ and $\delta \varepsilon_{01}$ have reasonable values.

From figure~\ref{Fig:quantum} we find that condition~(iii) is violated if $\kappa$ is tuned too close to $\kappa_c(a)$ whereas condition~(ii) is violated if $\kappa$ becomes too large. Energy splittings $\delta \varepsilon_{01}$ corresponding to approximately {\color{black}$ 25\units{mK}$} look promising. To check condition~(i) we calculate the energy difference $\Delta U$ from~\eqref{Eq:energy:diff} and find values corresponding to approximately between {\color{black}$ 100\units{mK}$ and $ 135\units{mK}$}. Therefore, we may expect to observe coherent quantum oscillations as defined by~\eqref{eq:P_L} with a frequency $\Delta_{01}/(2\pi)=0.52\units{GHz}$ for temperatures below {\color{black}$25\units{mK}$} which is {\color{black}experimentally accessible}.

\begin{figure}[!hbt]
  \centering
  \includegraphics{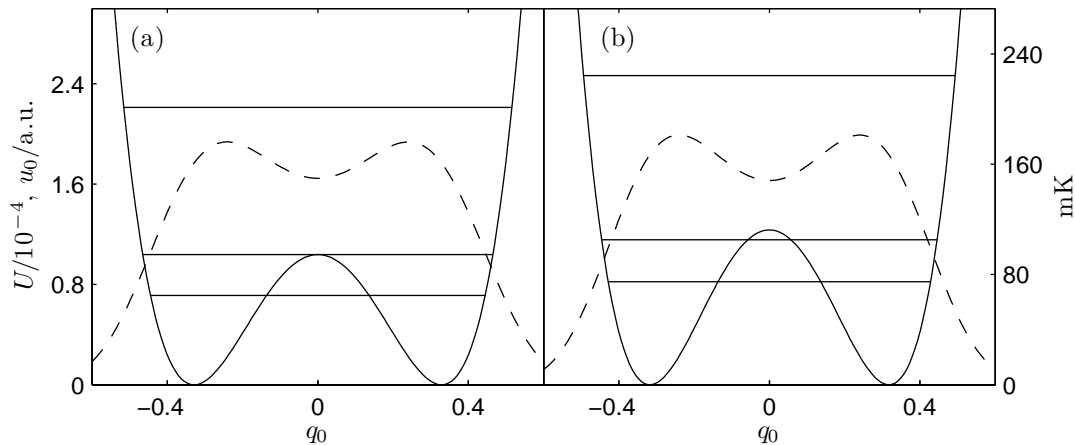}
  \caption{The potential energy $U$~\eqref{Eq:quartic} as a function of $q_0$ for $a=1.8$, $\kappa=0.08\pi$,  (a) and $a=1.6$, $\kappa=0.68\pi$ (b). The horizontal lines represent the corresponding eigenvalues $\varepsilon_0$, $\varepsilon_1$ and $\varepsilon_2$ of the Schr\"{o}dinger equation \eqref{Eq:Schroedinger:quartic}. The dashed lines depict the quantum-mechanical ground states $u_0(q_0)$ in arbitrary units. As in figure~\ref{Fig:quantum} the right vertical axes shows the temperatures corresponding to the energies of the left axis. In both cases the energy splitting $\delta \varepsilon_{01}$ corresponds to $25\units{mK}$.
}
  \label{Fig:potential}
\end{figure}

Two typical examples are shown in figure~\ref{Fig:potential}; one for $a=1.8$ and {\color{black}$\kappa=0.08\pi$} (a), and one for $a=1.6$ and $\kappa=0.68\pi$ (b), indicated by two arrows in figure~\ref{Fig:quantum}. The first example corresponds to the case $a>a_c(0)$ while the second example corresponds to the case $a_c(\pi)<a<a_c(0)$. The results for the two regimes look similar: For the parameters of figure~\ref{Fig:potential} (a) we find {\color{black}$\Delta U=1.23\times10^{-4}$ (111\units{mK})}, {\color{black}$\delta \varepsilon_{01}=0.27\times10^{-4}$ (25\units{mK})} and {\color{black}$\delta \varepsilon_{12}=1.19\times10^{-4}$ (109\units{mK})} while for the parameters of figure~\ref{Fig:potential} (b) we find {\color{black}$\Delta U=1.47\times10^{-4}$ (135\units{mK})}, {\color{black}$\delta \varepsilon_{01}=0.27\times10^{-4}$ (25\units{mK})} and {\color{black}$\delta \varepsilon_{12}=1.34\times10^{-4}$ (122\units{mK})}. From figure~\ref{Fig:quantum} we conclude that for larger values of $a$ the 
energy
splitting $\delta \varepsilon_{01}$ becomes more sensitive to $\kappa$. The advantages and disadvantages of both regimes in terms of read-out are discussed in the next section.

\section{Read-out}
\label{Sec:Read-out}

To observe coherent quantum oscillations one has to distinguish between the $(\kappa,\kappa-2\pi)$ state described by $\mu_+(x)$ and the $(\kappa-2\pi, \kappa)$ state described by $\mu_-(x)$ of the molecule. Therefore, it is necessary to readout its state. One possibility is to readout the flux associated with each fractional vortex in a molecule, similar to the earlier experiments~\cite{Dewes:2008:ReArrangeE}. The magnetic fluxes measured in these states should be distinguishable.

\begin{figure}[!hbt]
  \centering
  \includegraphics{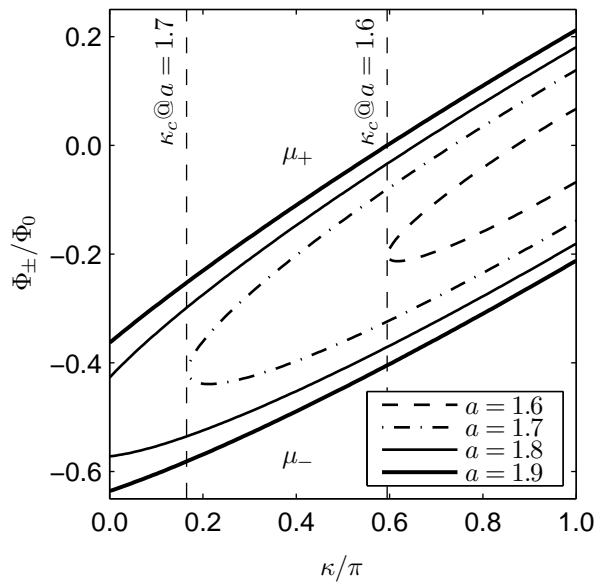}
  \caption{%
    The fluxes $\Phi_+$ and $\Phi_-$ defined by~\eqref{Eq:FluxL} as a function of $\kappa$ for different values of $a$. For each value of $a$ the upper branch corresponds to $\Phi_+$ and the lower one to $\Phi_-$.
  }
  \label{Fig:flux}
\end{figure}

In order to see if this is possible we calculate the flux
\begin{equation}
 \Phi_\pm \equiv \frac{\Phi_0}{2\pi}\int_{-\infty}^0\mu_{\pm,x}\, \dd x
 = \frac{\Phi_0}{2\pi}[\mu_\pm(0)-\mu_\pm(-\infty)]
 \label{Eq:FluxL}
\end{equation}
measured on the left half of the LJJ for the two different states $\mu_-(x)$ and $\mu_+(x)$ and the results are shown in figure~\ref{Fig:flux}.

For $a<a_c(0)$ the flux difference $\Delta \Phi \equiv \Phi_+-\Phi_-$ vanishes at the bifurcation points $\kappa_c(a)$. For $a>a_c(0)$ the value of $\Delta\Phi$ remains finite for $\kappa\to0$, as in this limit the two states correspond to two integer fluxons weakly pinned at $x=\pm a/2$. For larger values of $a$ the flux difference $\Delta \Phi$ increases. Therefore, in junctions with larger $a$ the two states are easier to distinguish. In this case, however, the quantum system is harder to control because the relevant range of $\kappa$ becomes smaller, see figure~\ref{Fig:quantum}.

\section{{{}}{Fluctuations}}
\label{Sec:Decoherence}

{We expect that two sources of noise will play a major role in our system. For the sake of simplicity we assume in our qualitative discussion that the fluctuations are quasistatic (low frequency noise).}

First, there are the fluctuations in the injector current circuitry. This noise was already identified as bottleneck in our previous studies. Therefore, we use current injectors in persistent mode in our latest experiments~\cite{kemmler:2010}. {Additionally, we have designed our setup such that the injector currents are relatively small at the working point, see section~\ref{Sec:Initializing}. The effect of {fluctuating injector currents} (assuming common noise in both injector pairs due to the persistent mode) can directly be seen from our formulas. It results in noise in $\kappa$ and therefore a noisy barrier height which is shown in figure~\ref{Fig:energy:tune} (a). From this figure we conclude that the system is less sensitive to fluctuations for {{}}{smaller values of $a$. In particular, the region $a_c(\pi)<a<a_c(0)$ is more favourable than the region $a>a_c(0)$. 

{Second, one may expect flux noise (a spurious external field) to be another major source of {{}}{fluctuations}. Two components of the field are relevant in this case: $H_y$ (in-plane) and $H_z$ (perpendicular to the plane of the structure). The noise in $H_y$ is irrelevant, as our LJJ is (formally infinitely) long and $H_y$ enters only in the boundary conditions at the edges. In essence, the AFM molecule is protected from the fluctuations in $H_y$ as $H_y$ is screened by the LJJ on the length $\sim\lambda_J$ from the edge. The vertical component $H_z$ is expelled by the screening currents in the electrodes, but refocuses as a non-uniform flux density~\cite{Monaco:2008:LJJ+VertMagField,Scharinger:2010:MJJ:Ic(H)} $B_y(x)$. For relatively long JJs the $H_z$ to $B_y$ refocusing 
factor can be rather large for wide bias electrodes. 
However, the bias leads 
can be made only as long as the {AFM molecule} ($\sim2\ldots5\lambda_J$) or to be absent at all. In fact our {system} operates without bias lines. {In addition}, the profile of $B_y(x)$ is such that it has zero derivative in the middle of the LJJ, \ie, where the {AFM molecule} is situated. }

{Thus, we conclude that the {system discussed here} can be made quite insensitive to flux noise. A separate in-depth investigation, which takes into account, \eg, spatially non-uniform noise or high frequency noise will be published elsewhere.}

\section{Conclusions}
\label{Sec:Conclusions}

{{}}{We have presented the concept of a macroscopic quantum system consisting of of two fractional Josephson vortices where coherent quantum oscillations can be observed.} Two degenerate ground states are separated by an energy barrier, which can be tuned during the experiment by changing simultaneously the values of the discontinuities. The concept may work in both linear and annular geometries.

In particular, we have obtained analytical solutions for the stationary phases and their energies in unbiased LJJs with two discontinuities. Furthermore, we have analysed the stability of the stationary solutions by calculating the corresponding eigenmodes numerically. We have used these eigenmodes to map the low-energy dynamics of the system to the dynamics of a particle in a one-dimensional double-well potential and have solved the corresponding Schr\"{o}dinger equation.

We have shown how the energy barrier can be tuned with the help of injector currents to reach the quantum limit. Our results indicate that for typical parameters a quantum-mechanical two-level system can be established for temperatures below {$25\units{mK}$ which is at the limit of modern dilution refrigerators.} 

{Finally, we have analysed the sensitivity of the system to the most obvious sources of fluctuations. In essence, the system can be designed to be quite insensitive to flux noise. In experiments most attention should be paid to provide noise-free injector currents. We have found that our quantum system is less sensitive to noise of the injector currents for small values of $a$.}}

Experiments with such fractional vortex molecules are in progress in the T\"{u}bingen group.

\ack

We thank L. T. Haag, M. Zimmermann, J. Fischbach, T. Gaber, K. Buckenmaier, and J. M. Meckbach for many fruitful and stimulating discussions. This work was supported by the German Science Foundation (DFG) within SFB/TRR21.

\appendix

\section{Elliptic integrals and elliptic functions}\label{app:a}

In this appendix we introduce the elliptic integral of first kind and the Jacobi amplitude function and provide a collection of formulas which are used in the present paper. These relations are in accordance with~\cite{nist2011}.

\subsection{Elliptic integrals}

The elliptic integral of first kind is defined by
\begin{eqnarray}
  \F(\phi,k)&\equiv&\int\limits_0^\phi \frac{\dd\theta}{\sqrt{1-k^2\sin^2\theta}} 
  , \label{Eq:app:ellip:int:first}
\end{eqnarray}
where the argument $k$ is called modulus. 

For $\phi=\pi/2$ it reduces to the complete elliptic integral of first kind
\begin{eqnarray}
  \K(k)&\equiv&\F(\pi/2,k) 
  . \label{Eq:app:ellip:int:first:complete}
\end{eqnarray}
The elliptic integral $\F(\phi,k)$ obeys the symmetry relation
\begin{eqnarray}
  \F(n\pi\pm\phi,k)&=&2n\K(k)\pm\F(\phi,k) 
  , \label{Eq:app:F:symmetry}
\end{eqnarray}
where $n$ is an integer.

\subsection{Jacobi amplitude function}

For $k<1$ the elliptic integral \eqref{Eq:app:ellip:int:first} is a monotonously increasing function of $\phi$. The Jacobi amplitude function is the inverse of the elliptic integral. For a given value $u\equiv \F(\phi,k)$ we have to find the corresponding integration limit $\phi$. Then the Jacobi amplitude function is defined by
\begin{equation}
 \am(u,k)\equiv\phi 
 . \label{Eq:app:am_F}
\end{equation}
The relation
\begin{equation}
 \am(u,k)\equiv \arcsin\{k^{-1}\sin[\am(ku,k^{-1})]\}
   \label{Eq:app:a:1:am}
\end{equation}
extends this definition for $k>1$.

The Jacobi amplitude function is monotonously increasing for $k<1$ and periodic for $k>1$. Furthermore, it solves the differential equation
\begin{equation}
 \left[\frac{\dd}{\dd u}\am(u,k)\right]^2=1-k^2\sin^2[\am(u,k)] 
  \label{Eq:app:a:dn:diff:2}
\end{equation}
and for $k\to 1$ it reduces to
\begin{eqnarray}
  \am(u,1)&=&2\arctan(\ee^{u})-\pi/2 
   . \label{Eq:app:k=1:am}
\end{eqnarray}

The three Jacobi elliptic functions
\begin{eqnarray}
  \sn(u,k)&\equiv&\sin[\am(u,k)] 
  , \label{Eq:app:a:0:sn}
\end{eqnarray}
and
\begin{eqnarray}
  \cn(u,k)&\equiv&\cos[\am(u,k)] 
  , \label{Eq:app:a:0:cn}
\end{eqnarray}
as well as
\begin{eqnarray}
  \dn(u,k)&\equiv&\frac{\dd}{\dd u}[\am(u,k)] 
  , \label{Eq:app:a:1:dn:diff:app}
\end{eqnarray}
together with the Jacobi epsilon function
\begin{eqnarray}
  \mathcal{E}(u,k)&\equiv&\int\limits_0^u\dd t\ \dn^2(t,k) 
   \label{Eq:app:epsilon:1}
\end{eqnarray}
are defined in terms of the Jacobi amplitude function $\am(u,k)$.

These functions satisfy the relations
\begin{eqnarray}
  \sn(u,k)&=& k^{-1}\sn(ku,k^{-1}) 
  , \label{Eq:app:a:1:sn}
\end{eqnarray}
and
\begin{eqnarray}
  \cn(u,k)&=& \dn(ku,k^{-1}) 
  , \label{Eq:app:a:1:cn}
\end{eqnarray}
as well as
\begin{eqnarray}
  \dn(u,k)&=& \cn(ku,k^{-1}) 
  , \label{Eq:app:a:1:dn}
\end{eqnarray}
and the addition theorem
\begin{eqnarray}
  \mathcal{E}(u_1&+&u_2,k)=\mathcal{E}(u_1,k)+\mathcal{E}(u_2,k)\nonumber\\
  &-&k^2\sn(u_1,k)\sn(u_2,k)\sn(u_1+u_2,k) 
  . \label{Eq:app:epsilon:add}
\end{eqnarray}

\section*{References}


\providecommand{\newblock}{}

\end{document}